\documentclass[review,jog]{igs}

\usepackage[utf8]{inputenc}

\usepackage{mathabx}
\usepackage{graphicx}
\usepackage{siunitx}
\usepackage{xcolor}
\usepackage{tabularx}
\usepackage{longtable}
\usepackage{pdflscape}
\usepackage{hyperref}

\usepackage{setspace}
\begin{document}
\nolinenumbers

\title[GlacierMIP Protocol]{The Glacier Model Intercomparison Project (GlacierMIP): Towards standardisation of global glacier projections}

\author[van der Laan, Hock, and others]{Larissa Nora van der Laan$^{1*}$,
Regine Hock$^{2,3*}$, 
Johannes Brunner$^2$,
Yeliz A. Yilmaz$^2$,
Claire Wilson$^{9}$,
Lilian Schuster$^{10}$, 
Lander Van Tricht$^{5,6,10}$,
Ruitang Yang$^2$,
Harry Zekollari$^{10,5}$,
Matthias Huss$^{5,6}$, 
Ben Marzeion$^7$,
Fabien Maussion$^8$, 
David R. Rounce$^9$,
Johannes J. Fürst$^4$}

\affiliation{%
$^1$Niels Bohr Institute, Copenhagen University, Copenhagen, Denmark\\
$^2$Department of Geosciences, University of Oslo, Oslo, Norway\\
$^3$University of Alaska Fairbanks, Fairbanks, USA\\
$^4$Institute of Geography, Friedrich-Alexander-Universität Erlangen-Nürnberg, Erlangen, Germany\\
$^5$Laboratory of Hydraulics, Hydrology and Glaciology (VAW), ETH Z\"urich, Z\"urich, Switzerland\\
$^6$Swiss Federal Institute for Forest, Snow and Landscape Research (WSL), Sion, Switzerland\\
$^7$Institute of Geography and MARUM - Center for Marine Environmental Sciences, University of Bremen, Bremen, Germany\\
$^8$School of Geographical Sciences, University of Bristol, Bristol, UK\\
$^9$Department of Civil and Environmental Engineering, College of Engineering, Carnegie Mellon University, Pittsburgh, PA, USA\\
$^{10}$Department of Water and Climate, Vrije Universiteit Brussel, Brussel, Belgium\\[6pt]

Correspondence: Regine Hock and Larissa Nora van der Laan 
\email{regineho@uio.no, larissa.vdlaan@nbi.ku.dk}}

\begin{frontmatter}

\maketitle
\textsuperscript{*}~\textit{Contributed equally to this work.}

\begin{abstract}
The Glacier Model Intercomparison Project (GlacierMIP) provides an international framework for the coordinated evaluation and comparison of global glacier evolution models and their projections of glacier change. Its fourth phase, GlacierMIP4, updates previous global glacier projections through a standardised experimental design. For the first time, the protocol includes simulations extending to 2300 and an overshoot scenario for a subset of climate projections and requires monthly output of individual mass-balance components such as melt, accumulation and refreezing, in addition to annual results for glacier mass and area.
While permitting model-specific approaches to calibration and other aspects of model implementation, the protocol prescribes common atmospheric forcing data, inventory boundary conditions, and reporting conventions to enable consistent analyses of inter-model differences. Here, we describe the GlacierMIP4 experimental protocol, including the simulation domain, initialisation, atmospheric forcing, and standardised output requirements. More broadly, we emphasize the need for standardised terminology, units, sign convention and file formats in global glacier modelling to improve the traceability, interoperability, and comparability of results, and recommend this protocol as a blueprint for future global glacier modelling studies.

\end{abstract}
\end{frontmatter}

\clearpage
\section{ Introduction}

The Glacier Model Intercomparison Project (GlacierMIP) is a Targeted Activity of the Climate and Cryosphere (CliC) project of the World Climate Research Project (WCRP).
Its overall goal is to provide a framework for coordinated intercomparisons of global-scale glacier evolution models and their decadal-to-centennial projections, to provide community-based projections, e.g., to inform international assessments such as the Intergovernmental Panel on Climate Change (IPCC), to foster model improvements and to reduce uncertainties in projections of global glacier change. GlacierMIP focuses on all glaciers outside the ice sheets in Greenland and Antarctica but includes those in their periphery.

GlacierMIP complements the Ice Sheet Model Intercomparison Project (ISMIP), an umbrella framework under which several ice-sheet model intercomparisons have been coordinated, for example, ISMIP-HOM \citep{pattyn_2008}, ISMIP-HEINO \citep{calov_2010} and most prominently ISMIP6 \cite[e.g.,][]{Seroussi_2020,Goelzer_2020_ISMIP,Nowicki_2016}. ISMIP was endorsed by the Coupled Model Intercomparison Project Phase 6 (CMIP6, \cite{Eyring_2016}.
Preceding ISMIP, the European Ice Sheet Modelling Initiative (EISMINT) conducted some of the first coordinated, idealized benchmark experiments for ice-sheet models during the 1990s, including experiments on ice dynamics and thermomechanical coupling \citep{Huybrechts_1996,Payne_2000}.

Global glacier modelling was hampered for a long time by the absence of a globally complete glacier inventory, necessitating upscaling to achieve globally aggregated results \citep{Radic_2010}. Consequently, only few glacier models, capable of supplying global-scale projections \citep{Raper_2006,Radic_2011} had been developed prior to the release of the first version of the Randolph Glacier Inventory \cite[RGI, ][]{Pfeffer_2014}. This inventory was the first to provide a glacier outline for almost all glaciers in the world \citep{Arendt_2012}. The RGI subsequently spurred the development of several global glacier models and large-scale projections within just a few years \citep{Marzeion_2012,Slangen_2012, Hirabayashi_2013,Giesen_2013,Radic_2013}.
As the global glacier-modelling community expanded and the number of available models increased, coordinated international model intercomparisons became feasible and timely. GlacierMIP was therefore launched in 2015.

Its first phase (2015-2019) focused on a comparison of the glacier volume and area projections 2015-2100 from six previously published modelling studies including 214 global model runs based on four emission scenarios \citep{Hock_2019_JGlac}. Because of this post-hoc approach, there was no standardisation of initial and boundary conditions. The results were used in the Special Report on the Ocean and Cryosphere in a Changing Climate (SROCC) by the Intergovernmental Panel on Climate Change (IPCC) \citep{Hock_2019_SROCC}.

In GlacierMIP2 (2019-2020), all participating modelling groups were required to submit annual glacier volume and area projections until 2100, based on standardised experiments. Models had to cover at least one of the 19 primary RGI regions, and use projections from ten prescribed General Circulation Models (GCMs) forced by at least two emission scenarios (Representative Concentration Pathway RCP2.6 and RCP8.5 \citep{vanVuren_2011}. Regional initial ice volumes were prescribed based on RGI 6.0 \citep{huss_2012_thickness}. Results based on an ensemble of 288 unique projections from 11 glacier models were used to partition the total uncertainty into individual contributions caused by glacier models, GCMs, RCPs, and natural variability \citep{Marzeion_2020}. Results also informed the glacier projections used in the IPCC Sixth Assessment Report \citep{Fox-Kemper_2021}.

GlacierMIP3 (2021–2025) differed from the previous phases by aiming to quantify steady-state glacier volume and area under stabilized global mean temperature levels (e.g., +1.5$^\circ\mathrm{C}$, +2$^\circ\mathrm{C}$ relative to pre-industrial conditions), as well as the time required for glaciers to reach their new steady-state. The stabilisation experiments were based on eight glacier models forced with 80 prescribed constant climatic scenarios derived from five GCMs, and were run for 2000 to 5000 years \citep{Zekollari_Schuster_2025}.

The current phase, GlacierMIP4, aims to update global-scale glacier projections using the latest generation of global glacier evolution models and to investigate inter-model differences by examining individual components of simulated mass balance. Thus, for the first time, the GlacierMIP protocol requires the submission of individual mass-balance components (e.g. accumulation, melt, runoff) in addition to annual glacier mass and area. GlacierMIP4 also differs from the previous phases by requesting monthly output and simulations extending to 2300 for a subset of GCM-SSP combinations. 

Here, we describe the detailed GlacierMIP4 experimental protocol, including the model domain, simulation period, climate forcing, inventory boundary conditions, and output requirements. We also discuss and emphasize the need for improved standardisation of global glacier-model output as the global glacier-modelling community continues to grow. We outline how the GlacierMIP4 protocol can be adopted more broadly as a framework for future global glacier-modelling studies.


\section{ Experimental Design}

The GlacierMIP4 protocol prescribes key elements, including a common model domain, simulation period, climate forcing, inventory data, and set of output variables, while allowing for some flexibility in other aspects. This approach balances inter-model comparability with model-specific constraints, acknowledging the diversity of available models. Existing global glacier evolution models differ in their model physics (e.g., the treatment of glacier surface mass balance, frontal ablation, glacier dynamics and resulting geometric adjustments) as well as in their underlying calibration techniques. They also vary in spatial discretisation (elevation bands vs. grids), atmospheric forcing (only temperature and precipitation, or additional variables for an energy-balance approach), temporal resolution (monthly or daily) and initialisation procedures (e.g., with or without spin-up) \cite[see summaries in][]{Hock_2019_JGlac,Marzeion_2020,Zekollari_Schuster_2025}.

\subsection{ Simulation Period} 
Simulations are divided into hindcasts (calendar years 2000 through 2025) and projections, which begin in 2026 and extend through 2100 or 2299/2300, depending on the temporal coverage of the atmospheric forcing data. Hindcast simulations begin in 2000 because the RGI 7.0 has a nominal reference date of 2000, although the actual dates of individual glacier outlines may differ substantially from this target date \citep{RGIConsortium2023}. The projections are initialised using the glacier geometries (volume, area, surface elevation) simulated by the individual models at the end of the hindcast period.

\begin{table}
\caption{Regional glacier area and volume prescribed in GlacierMIP4. Areas are from RGI 7.0 \citep{RGIConsortium2023} and volumes are from \cite{Maffezzoli_2026}.}
\label{tab:area_volume}
\begin{tabular}{l l S[table-format=6.0] S[table-format=2.2]}
\hline
ID & RGI region name & {Area (km$^2$)} & {Volume ($10^3$ km$^3$)} \\
\hline
01 & Alaska                              & 86708  & 16.9  \\
02 & Western Canada and USA              & 14521  & 1.5   \\
03 & Arctic Canada North                 & 105370 & 24.3  \\
04 & Arctic Canada South                 & 40538  & 7.1   \\
05 & Greenland Periphery                 & 90482  & 13.2  \\
06 & Iceland                             & 11060  & 4.6   \\
07 & Svalbard and Jan Mayen              & 33959  & 6.7   \\
08 & Scandinavia                         & 2948   & 0.35  \\
09 & Russian Arctic                      & 51595  & 12.8  \\
10 & North Asia                          & 2643   & 0.19  \\
11 & Central Europe                      & 2124   & 0.11  \\
12 & Caucasus and Middle East            & 1407   & 0.08  \\
13 & Central Asia                        & 50344  & 3.8   \\
14 & South Asia West                     & 33075  & 3.8   \\
15 & South Asia East                     & 16049  & 1.0   \\
16 & Low Latitudes                       & 1929   & 0.09  \\
17 & Southern Andes                      & 27674  & 6.7   \\
18 & New Zealand                         & 886    & 0.08  \\
19 & Subantarctic and Antarctic Islands  & 133432 & 45.7  \\
\hline
   & Global                              & 706744 & 149   \\
\hline
\end{tabular}
\end{table}

\subsection{ Domain and Inventory Data} 
As in previous GlacierMIPs, participants use the RGI to define standardised model domains and glacier outlines. While participants are encouraged to provide data for all 19 primary RGI regions \citep{RGIConsortium2023}, the minimum requirement for participation is the simulation of one entire RGI region. This approach aims to balance the desire for broad regional coverage with a sufficiently low threshold to encourage wide international participation, and thus model diversity. The protocol prescribes the use of RGI 7.0 \citep{RGIConsortium2023,Maussion2023} for glacier area and outlines, which includes approximately 275,000 glaciers covering an area of 707,000 km$^2$.

GlacierMIP4 uses the IceBoost v2.0 by \cite{Maffezzoli_2026} as its ice thickness reference. The dataset provides gridded ice thickness at the individual and glacier complex level with an approximate reference period of 2010–2018 but no glacier-specific reference dates. We selected this dataset because other globally complete datasets currently available refer to RGI 6.0 rather than RGI 7.0 \citep{Farinotti_2019,Millan_2022,Frank_2026}. IceBoost v2.0 differs from previous global products by using a machine-learning approach trained on more than seven million ice-thickness observations, rather than relying primarily on physics-based ice-thickness inversions. 

As the timestamps of the glacier outlines differ notably, and exact dates for the ice thickness data are not available, participants are free either to treat thickness values as representative for the year 2000 and initialise their hindcast simulations accordingly, or to simulate them internally, for example through a dynamic spin-up or other forms of relaxation/initialisation strategies. However, for each region, the total volume is not allowed to differ by more than 10\% from the regional IceBoost v2.0 ice volumes (Table \ref{tab:area_volume}). The range accounts for the uncertainty in ice thickness dates and estimates and allows for inter-model differences in thickness initialization.

\subsection{ Atmospheric Forcing Data}

To facilitate direct comparison among glacier models, GlacierMIP4 prescribes common climate forcing datasets for all participants. For the hindcast period (2000-2025), all glacier models are required to use ERA5, the fifth-generation atmospheric reanalysis produced by the European Centre for Medium-Range Weather Forecasts (ECMWF)
\citep{Sociera5}.
For the projection period, starting in 2026, participants are required to use bias-corrected output from the Coupled Model Intercomparison Project Phase 6 (CMIP6) \citep{Eyring_2016}, under several Shared Socio-economic Pathways (SSPs) listed in Table \ref{tab:gcms}.

A total of eight coupled General Circulation Models or Earth System Models (for simplicity henceforth referred to as GCMs) were selected to provide a compact but diverse ensemble of climate forcing, while keeping the number of simulations computationally feasible. The selection is based primarily on the uncertainty-informed model subset framework by \citet{Snyder_2024}, which has been developed to identify a small subset of CMIP6 models that retains important characteristics of the broader ensemble. Their approach jointly considers temperature and precipitation projections and uses the first five principal components of the resulting climate response space to identify subsets that preserve the characteristics of ensemble variability. In addition, candidate subsets for GlacierMIP4 are required to roughly preserve the 'very likely' (5-95\%) range of equilibrium climate sensitivity (ECS) of 2$^\circ\mathrm{C}$ - 5$^\circ\mathrm{C}$ assessed by the IPCC \citep{IPCC_2021_Ch_7_ECS}. From \cite{Snyder_2024}´s Table 3, GlacierMIP4 selects all five models from Experiment 2 (IPSL-CM6A-LR, ACCESS-ESM1-5, MRI-ESM2-0, MPI-ESM1-2-HR, and MIROC6) and BCC-CSM2-MR from Experiment 1. Experiment 2 explicitly accounts for structural model dependence by excluding the models with clear dependencies on other models before constructing the subset \citep{Snyder_2024}. This selection therefore provides a targeted sample of CMIP6 climate response rather than a simple selection based on climate sensitivity alone.

The GCM selection has additionally been designed to align closely with the Ice Sheet Model Intercomparison Project (ISMIP7, \url{https://www.ismip.org/}, successor to ISMIP6), which uses CESM2-WACCM and MRI-ESM2-0 for its core experiments. MRI-ESM2-0 is already included in the \cite{Snyder_2024} subset, while CESM2-WACCM provides an additional model that is not included in their Experiment 2 subset because of its structural dependence on other CMIP6 models. NorESM2-MM is also included in GlacierMIP4 because it represents a comparatively low sensitivity (ECS = 2.49~K) counterpart to CESM2-WACCM (ECS = 4.68~K, Table \ref{tab:gcms}). NorESM2-MM and CESM2-WACCM are distinct models but share part of the broader CESM model heritage. Their inclusion therefore does not maximize structural independence within the ensemble. This illustrates the trade-off between minimizing model dependence and sampling a broad range of plausible climate responses \citep{Dunne_2025}. The eight selected GCMs span the 11$^{th}$ to 73$^{rd}$ percentiles of the distribution of ECS values across the 55 CMIP6 climate models investigated by \cite{Hausfather_2022}. Across these 55 models, the median ECS is $3.76\,^\circ\text{C}$ and the interquartile range is $2.97-4.74\,^\circ\text{C}$.
For each GCM, only the r1i1p1f1 realisation is used.

The SSPs are chosen to sample a broad range of plausible climate pathways (SSP1-2.6, SSP3-7.0, SSP5-8.5, and overshoot scenario SSP5-3.4-OS). In the latter scenario, emissions initially follow the high-emissions SSP5-8.5 pathway until 2040, then decline extremely rapidly. It is added to assess the extent to which glacier mass loss can be reversed following peak warming. We note that some SSP1-2.6–forced GCM projections 
also exhibit a reversal in warming trend, with peak warming in the late 21st century or shortly thereafter. The prescribed GCM-SSP combinations sample both uncertainty in the climate model response and uncertainty associated with the future forcing trajectories.

In total, the protocol requires each glacier model to perform simulations for 28 distinct GCM–SSP combinations (eight GCMs paired with three or four SSPs, depending on the GCM), of which eight extend to 2300 and three (using CESM2-WACCM) to 2299 (Table \ref{tab:gcms}).

\begin{figure}
        \centering
        \includegraphics[width=0.8\linewidth]{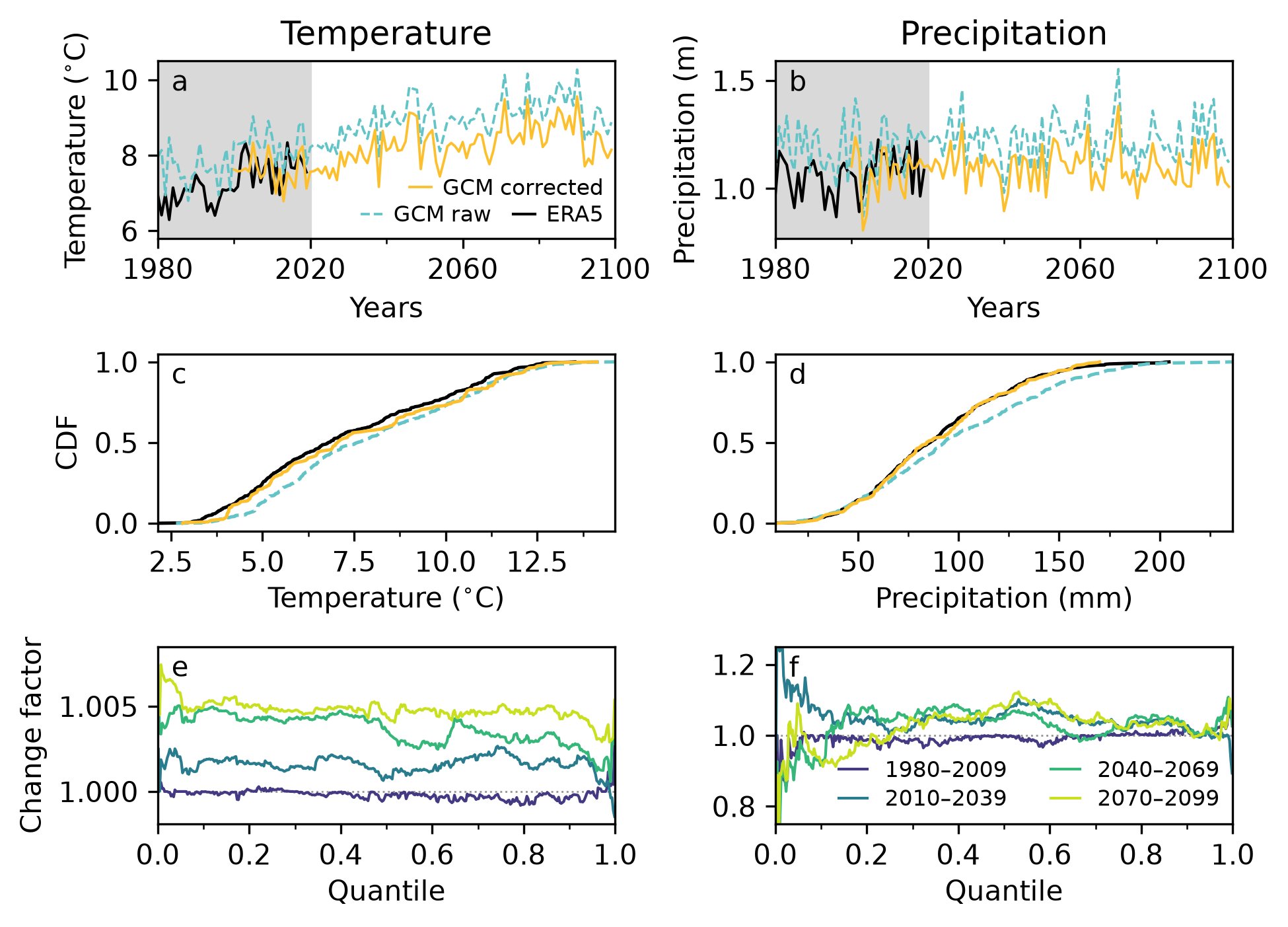}
        \caption{Illustration of the quantile delta mapping method to downscale the GCM projection relative to the ERA5 data. (a-b) Annual mean temperature and total precipitation from ERA5 and the raw and bias-corrected GCM data. (c-d) Corresponding empirical cumulative distribution function (CDF). (e-f) Corresponding change factor for four different periods. GCM data are from CESM2-WACCM under SSP3-7.0 and the closest grid cell to Gulkana Glacier, Alaska. The shaded area marks the ERA5 reference period.}
        \label{fig:downscaling}
\end{figure}

As part of GlacierMIP4, monthly temperature and precipitation of all GCM projections are bias-corrected to the ERA5 reanalysis to ensure that all participants use consistent climate-forcing data. 
Bias-correction of temperature and precipitation data (Fig.~\ref{fig:downscaling}a,b) was performed using the quantile delta mapping method \cite[e.g.,][]{Cannon_2015,Lader_2017}. GCM data are first resampled onto the ERA5 grid using nearest-neighbor interpolation. 
Cumulative distribution functions are then computed for each grid cell for both the coarse GCM historical output and the finer-resolution ERA5 reference data over the shared historical period 1980-2019, and for consecutive 30-year windows over the full GCM period (1980-2009, 2010-2039, 2040-2069, 2070-2099, etc.) (Fig.~\ref{fig:downscaling}c,d). Within each 30-year window, each monthly temperature/precipitation value is assigned its quantile within that window's distribution. The change factor between this value and the value at the same quantile of the GCM's 1980-2019 distribution is computed (Fig.~\ref{fig:downscaling}e,f). This change factor is applied to the value at the same quantile of the ERA5 1980-2019 distribution. The corrected data then match the statistics of ERA5 within the reference period, while preserving the GCM-projected change at each quantile. 
For monthly precipitation below 0.001 m day$^{-1}$, an additive correction is used to avoid unbounded change factors \cite[]{Lader_2017}, and corrected values are clipped to be non-negative.

Models requiring variables in addition to temperature and precipitation (e.g. radiation components), or a different temporal resolution (e.g., daily), are required to prepare and bias-correct the GCM data independently.

\begin{table}
\caption{List of the eight prescribed CMIP6 General Circulation Models or Earth System Models (GCMs) and four Shared Socioeconomic Pathway (SSP) emission scenarios used to provide atmospheric forcing for the glacier models. $X$ indicates available projections and stars ($*$) indicate GCMs with simulations until 2299/2300. ECS is the equilibrium climate sensitivity (in K). The percentile column indicates the CMIP6 ECS percentiles which shows the position of the eight selected GCM within the distribution of their 55 models' ECS values.}
\label{tab:gcms}
\begin{tabular}{l S[table-format=1.2] c c c c c}
\hline
GCM & {\text{ECS\textsuperscript{a}}} & {\text{Percentile}} & SSP1-2.6 & SSP3-7.0 & SSP5-8.5 & SSP5-3.4OS \\
\hline
ACCESS-ESM1-5  & 3.88 & 56\% & X$^*$ & X & X$^*$ &      \\
BCC-CSM2-MR    & 3.02 & 31\% & X\phantom{$^{*}$}     & X & X\phantom{$^{*}$}     &      \\
CESM2-WACCM    & 4.68 & 71\% & X$^*$ & X & X$^*$ & X$^*$ \\
IPSL-CM6A-LR   & 4.70 & 73\% & X$^*$ & X & X$^*$ & X$^*$ \\
MRI-ESM2-0     & 3.13 & 38\% & X$^*$ & X & X$^*$ & X$^*$ \\
MPI-ESM1-2-HR  & 2.98 & 27\% & X\phantom{$^{*}$}     & X & X\phantom{$^{*}$}     &      \\
MIROC6         & 2.60 & 16\% & X\phantom{$^{*}$}     & X & X\phantom{$^{*}$}     & X\phantom{$^{*}$}    \\
NorESM2-MM     & 2.49 & 11\% & X\phantom{$^{*}$}     & X & X\phantom{$^{*}$}     &      \\
\hline
\end{tabular}

\vspace{4pt}
{\footnotesize
\textsuperscript{a}ECS is defined as "the equilibrium value of global surface air temperature change in response to a sustained doubling of atmospheric $\text{CO}_2$ concentration from a pre-industrial reference state" \citep{IPCC_2021_Ch_7_ECS}. Values taken from the column 'ECS150' in Supplementary Data in \cite{Hausfather_2022}
}
\end{table}

\subsection{ Calibration and Validation}

Due to differences in underlying physics and implementation, each glacier model has its own set of parameters. These typically include parameters that relate climate variables to snow accumulation and snow/ice melt, and parameters that correct for biases in and downscale atmospheric forcing data. All of these parameters must be calibrated using available observations.

Model calibration and validation procedures and data are left to each participant's own discretion. This allows for the incorporation of individual preferences in terms of data and methods.
The protocol provides access to an RGI7.0-updated version of the glacier-wide specific geodetic mass-balance dataset by \cite{Hugonnet_2021}. Other suggested data types are glacier-wide mass-balance data from the World Glacier Monitoring Service \citep{WGMS2026}, snowline data \cite[e.g.,][]{Loibl_2025_snowline} and Gravity Recovery and Climate Experiment (GRACE) data \cite[e.g.,][]{Wouters_2019}.


\begin{table}
\caption{Summary of prescribed conventions, variable names, units, and
  metadata standards for GlacierMIP4.}
\label{tab:prescriptions}
\begin{tabular}{l l}
\hline
\textbf{Item} & \textbf{Prescribed standard} \\
\hline
\multicolumn{2}{l}{\textit{Domain and time}} \\
Spatial domain           & RGI7.0 regions (01-19) \\
Hindcast period          & 2000-2025 \\
Projection period        & 2026-2100 (2026-2299/2300 for selected GCM-SSP combinations) \\
Time unit                & days since 1850-01-01 \\
Time reference point      & Start of calendar year (annual), start of month (monthly) \\
\hline
\multicolumn{2}{l}{\textit{Inventory and identifiers}} \\
Glacier outlines/area    & RGI7.0 \\
Ice thickness            & \cite{Maffezzoli_2026} \\
Glacier identifier       & RGIId (String, individual-glacier submissions only) \\
\hline
\multicolumn{2}{l}{\textit{Parameter values}} \\
Ice density (mass$\to$volume conversion) & 900~kg~m$^{-3}$ \\
\hline
\multicolumn{2}{l}{\textit{Sign convention}} \\
Mass-gain terms (accumulation, refreezing) & Positive \\
Mass-loss terms (melt, frontal ablation) & Positive \\
\hline
\multicolumn{2}{l}{\textit{Metadata standard}} \\
Convention                & CF (Climate and Forecasting) metadata conventions \\
Temporal aggregation      & Explicit via \texttt{cell\_methods} and time-boundary variables \\
\hline

\multicolumn{2}{l}{\textit{File naming}} \\
Regional file             & [model]\_rgi[NN]\_[GCM or ERA5]\_[SSP]\_[annual/monthly].nc \\
Individual-glacier file   & As above, with 'indiv' appended to region code \\
\hline
\end{tabular}
\end{table}

\subsection{ Output Variables}

The GlacierMIP4 protocol stipulates the submission of the variables listed in Tables \ref{tab:state_vars} and \ref{tab:output_change_vars}, provided that these variables are applicable to their models. For example, not all models simulate frontal ablation or refreezing; in such cases, missing value should be given. Variables are reported at annual or monthly resolution, depending on the variable, and are provided as regional aggregates for each simulated RGI region.

In addition, participants are encouraged, where feasible, to provide the same output variables at the individual-glacier level. This finer spatial resolution enables a more detailed assessment of inter-model differences and their underlying causes, which can be difficult to disentangle from regionally aggregated output alone. It also allows evaluating spatial variations in glacier mass changes within an RGI region.

Output variables consist of state variables, reported at the beginning of each calendar year (Table \ref{tab:state_vars}) and time-integrated variables, reported over defined periods (monthly or annually, Table \ref{tab:output_change_vars}).
State variables include glacier area, glacier mass, and glacier mass below sea level of marine-terminating glaciers. The latter allows quantifying the sea-level contribution for glaciers, among others accounting for the displacement of ocean water by glacier ice below sea level when converting mass changes to sea-level equivalent. Time-integrated variables include mass-balance variables and glacier-model-corrected atmospheric forcing values.
The mass-balance variables comprise accumulation (excluding refreezing), melt, refreezing, glacier runoff and frontal ablation. 
Accumulation, melt and refreezing are calculated over the evolving glacierised area. Glacier runoff is defined as melt minus refreezing plus liquid precipitation, and likewise calculated over the evolving glacierised area \cite['moving-gauge' runoff, e.g.,][]{Bliss_2014} rather than over the initial glacierised area \cite['fixed-gauge' runoff, e.g.,][]{huss2018global,rounce2020_HMA,Rounce_2025_global-hydro}.
In contrast to the other mass-balance components, GlacierMIP4 requests annual rather than monthly output for frontal ablation because most current-generation global models, if they compute it at all, only do so at annual resolution.

Due to the coarse resolution of the ERA5 and bias-corrected GCM data, the glacier models often adjust the atmospheric input prior to computing the mass balance \cite[see summaries in][]{Hock_2019_JGlac,Marzeion_2020,Zekollari_Schuster_2025}. For example, models apply a glacier-specific temperature bias correction and a precipitation correction factor to the gridded forcing data. In addition, a temperature lapse rate and precipitation gradient may be used to distribute the temperature and precipitation across elevations. 
Thus, the actual temperature values used to calculate melt and to separate solid from liquid precipitation, as well as the precipitation values used in the accumulation calculations, may differ from the original input data and may vary among glacier models. 

Therefore, the protocol requests output of the temperature and precipitation values after any corrections to the atmospheric forcing data applied within the glacier models. The purpose of these additional output variables is to evaluate the extent to which inter-model differences in glacier response arise from differences in the air temperature and precipitation used in each model’s mass-balance calculations following model-specific adjustments.

Corrected temperature values are spatially averaged, and precipitation is summed, over the initial glacierised area (at start of the hindcast) of each simulated RGI region or individual glacier.
Temperature values must be area-weighted based on elevation-dependent or grid-based simulations, so that for each spatial unit (elevation band or grid cell), the temperature is weighted by that unit's initial area. 
Although forcing in newly de-glaciated areas is not included in the mass-balance calculation, using a constant initial glacierised area for spatial aggregation prevents the corrected forcing values from differing among glacier models because of changes in glacier area.

In addition to the variables in Tables \ref{tab:state_vars} and \ref{tab:output_change_vars},
GlacierMIP4 participants submitting individual glacier output are also encouraged to submit the variables in Table \ref{tab:optional_vars}. Annual and transient equilibrium line altitudes, and the accumulation area ratio provide useful indicators for the "health" of a glacier and enable comparison to available observations. Basin runoff is defined as melt minus refreezing plus liquid precipitation over the initial glacier area (´fixed-gauge' glacier runoff). In contrast to the ´moving-gauge' glacier runoff requested for the regionally-aggregated output (Table \ref{tab:output_change_vars}), it includes rain and melt from ice-free areas once the glacier has retreated. This variable may support supplementary analyses of glacier hydrology, and has been used as input to large-scale hydrological models \cite[e.g.,][]{Rounce_2025_global-hydro}.

\subsection{ Output Specifications}

The GlacierMIP4 protocol follows the NetCDF Climate and Forecasting (CF) conventions for output metadata \citep{CFmetadata} with respect to the temporal properties of the data. Each time step is associated with an interval defined by a time-bound variable. The temporal aggregation of each variable is specified in the cell methods attribute (e.g., "time: sum", "time: mean", "time:point") instead of being implied by the variable name. Time is defined consistently for annual and monthly output as days since 1 January 1850, referenced to the start of the corresponding year or month. 

All variable definitions, units, sign conventions, and metadata are summarized in Tables \ref{tab:prescriptions}-\ref{tab:optional_vars}.
For submissions of individual glacier output, glaciers are identified via the RGI7.0 glacier identifier (RGIId). All individual mass-balance components, including the mass-loss terms melt and frontal ablation, are defined as positive quantities. 

The protocol requires separate files for monthly and annual variables, for both regionally aggregated and optional individual-glacier output. For each simulated RGI region, participants are requested to submit one file at monthly and one file at annual resolution for the ERA5-forced hindcast (2000–2025) and for each of the 28 GCM–SSP scenarios for the projection period (2026–2100, extended to 2299/2300 for a subset of GCMs; Table \ref{tab:gcms}), thus a total of 58 files.

Output files follow a standardised naming convention that encodes the glacier model, RGI region, forcing dataset (ERA5 for the hindcast, or GCM and SSP for projections), and temporal resolution (Table \ref{tab:prescriptions}). The full specification and NetCDF templates are provided in the accompanying protocol repository \url{(https://github.com/GlacierMIP/GlacierMIP4)}.

\begin{table*}
\caption{Requested state variable output for GlacierMIP4 submissions, reported at the start of the year. The data are spatially aggregated over each simulated RGI region or provided at the individual glacier level.}
\label{tab:state_vars}
\begin{tabular}{l l l l l l}
\hline
\textbf{Variable} & \textbf{Short name} & \textbf{Unit} & \textbf{Data type} & \textbf{Cell method\textsuperscript{b}} & \textbf{Temporal res.} \\
\hline
Area                  & \texttt{area}        & m$^2$            & float32 & point & annual \\
Mass                 & \texttt{mass}        & kg               & float32 & point & annual \\
Mass below sea level  & \texttt{mass\_bsl}   & kg               & float32 & point & annual \\
Time (nc-axis)      & \texttt{time}        & days since 1850-01-01  & long    & ---   & annual \\
ID (nc-axis, optional)\textsuperscript{a}& \texttt{RGIId}       & ---              & String  & ---   & --- \\
\hline
\end{tabular}

\vspace{4pt}
{\footnotesize
\textsuperscript{a}ID of glacier in the Randolph Glacier Inventory (RGI 7.0). Only for individual glacier submissions.\\
\textsuperscript{b}Standard string attribute defined in the CF metadata convention to describe the method used to derive the data value over a specific spatial or temporal dimension (e.g., point, mean, maximum, or sum over an interval), here referring to time.}
\end{table*}

\begin{table*}
\caption{Requested mass-balance and forcing output variables for GlacierMIP4 submissions. The data are spatially aggregated over each simulated RGI region or provided at the individual glacier level.}
\label{tab:output_change_vars}
\begin{tabular}{l l l l l l}
\hline
\textbf{Variable} & \textbf{Short name} & \textbf{Unit} & \textbf{Data type} & \textbf{Cell method\textsuperscript{j}} & \textbf{Temporal res.} \\
\hline
Accumulation\textsuperscript{a}   & \texttt{acc}         & kg              & float32 & sum  & monthly \\
Melt\textsuperscript{b}           & \texttt{melt}        & kg              & float32 & sum  & monthly \\
Refreezing\textsuperscript{c}     & \texttt{refreeze}    & kg              & float32 & sum  & monthly \\
Glacier runoff\textsuperscript{d}         & \texttt{runoff\_glac}& kg              & float32 & sum  & monthly \\
Frontal ablation\textsuperscript{e}      & \texttt{frontal\_abl}& kg               & float32 & sum   & annual \\
Precipitation\textsuperscript{f}  & \texttt{precip}      & kg              & float32 & sum  & monthly \\
Temperature\textsuperscript{g}    & \texttt{temp}        & K               & float32 & mean & monthly \\
Time (nc-axis)\textsuperscript{h} & \texttt{time}        & days since 1850-01-01 & long    & ---  & monthly \\
ID (nc-axis, optional)\textsuperscript{i} & \texttt{RGIId} & ---            & String  & ---  & --- \\
\hline
\end{tabular}

\vspace{4pt}
{\footnotesize
\textsuperscript{a}Mass added to the glacier, excluding refreezing.\\
\textsuperscript{b}Mass removed from the glacier, excluding frontal ablation (e.g., melt, sublimation).\\
\textsuperscript{c}Total refreezing within the snowpack or glacier. Missing value if not simulated.\\
\textsuperscript{d}Runoff defined as melt minus refreezing plus liquid precipitation over the evolving glacierised area. The values represent 'moving-gauge' glacier runoff.\\
\textsuperscript{e}Mass loss at marine- and lake-terminating glaciers due to calving, frontal melting and sublimation above the waterline and sub-aqueous frontal melting below the waterline. Missing value if not simulated.\\
\textsuperscript{f}Total precipitation over the initial (start of the hindcast in 2000) glacierised area after any glacier-model-specific corrections to the ERA5 or GCM forcing data.\\
\textsuperscript{g}Mean near-surface (2~m) temperature area-weighted over the initial glacierised area (start of the hindcast in 2000) after any model-specific corrections to the ERA5 or GCM forcing data. 
\\
\textsuperscript{h}Start of the month or year.\\
\textsuperscript{i}ID of glacier in the Randolph Glacier Inventory (RGI 7.0). Only for individual glacier submissions.\\
\textsuperscript{j}CF metadata convention attribute defined in Table \ref{tab:state_vars}}
\end{table*}


\begin{table*}
\caption{Optional output variables for individual-glacier submissions.}
\label{tab:optional_vars}
\begin{tabular}{l l l l l l}
\hline
\textbf{Variable} & \textbf{Short name} & \textbf{Unit} & \textbf{Data type} & \textbf{Cell method\textsuperscript{e}} & \textbf{Temporal res.} \\
\hline
Basin runoff\textsuperscript{a}                      & \texttt{runoff\_basin} & kg       & float32 & sum     & monthly \\
Equilibrium line altitude\textsuperscript{b}         & \texttt{ELA}           & m a.s.l. & float32 & maximum & annual \\
Transient equilibrium line altitude\textsuperscript{c} & \texttt{snowline}    & m a.s.l. & float32 & maximum & monthly \\
Accumulation Area Ratio\textsuperscript{d}           & \texttt{AAR}           & ---      & float32 & minimum & annual \\
\hline
\end{tabular}

\vspace{4pt}
{\footnotesize
\textsuperscript{a} Runoff, defined as melt minus refreezing plus liquid precipitation over the initial glacier area. The values represent 'fixed-gauge' glacier runoff.\\
\textsuperscript{b}Elevation of the annual equilibrium line, defined as the highest elevation over the course of the calendar year where the climatic mass balance is zero, as derived from the mass-balance profile, i.e.\ the elevation where the profile intersects zero climatic mass balance.\\
\textsuperscript{c}Same as\textsuperscript{b}, but calculated for each month. In case of sub-monthly model resolution, this is the maximum value during the month.\\
\textsuperscript{d}Minimum ratio of accumulation area to total glacier area over the course of a year. The accumulation (ablation) area is the area where the annual mass balance is positive (negative).\\
\textsuperscript{e}CF metadata convention attribute defined in Table \ref{tab:state_vars}}
\end{table*}

\clearpage

\section{ Discussion and Conclusions}

\subsection{ From Patchwork to Standardisation}

Global glacier models have historically adopted their own output conventions, typically documented in model-specific publications and source-code repositories. However, these conventions vary widely because no commonly accepted standard exists. Previous GlacierMIP phases requested only projections of annual glacier volume and glacier area. The corresponding output formats were not widely adopted beyond the GlacierMIP exercises themselves. 

Divergence in model output standards occurs on various levels, ranging from differences in unit and variable naming to differences in terminology. To illustrate how this divergence manifests outside GlacierMIP, we compile the conventions of three widely used global glacier evolution models in Supplementary Table S1: 
the Open Global Glacier Model \cite[OGGM v.1.6.2,][]{Maussion2024_OGGM},
the Python Glacier Evolution Model \cite[PyGEM v.1.1.0,][]{Rounce_2023, tober_2026_projections} and 
the Global Glacier Evolution Model\cite[GloGEM v.0.1.1,][]{Huss_2015,Huss2026_GloGEM}. For example, PyGEM and OGGM report glacier area in m$^2$, while GloGEM reports it in km$^2$. PyGEM reports glacier mass in kg while OGGM and GloGEM express it as ice volume, in m$^3$ and km$^3$, respectively. In PyGEM mass changes are reported as volumes in m$^3$ water equivalent (w.e.), while the other two models use specific units (m w.e. yr$^{-1}$).
Converting between the two units requires information on the density of ice used, however, this factor is not always explicitly stated in a model's output metadata.
Variable naming conventions can also be confusing or misleading. For example, PyGEM uses the term "precipitation" to refer exclusively to “liquid precipitation,” whereas GloGEM more intuitively incorporates the term “rain” into the variable name.

At present, comparing previously published projections from different models requires a case-by-case review of each model's documentation to identify methodological choices, underlying assumptions and output conventions, a burden that a shared metadata standard could substantially reduce.

\subsection{ Recommendations Beyond GlacierMIP4}

Although our protocol was developed specifically for GlacierMIP4, it provides a practical blueprint for the glacier modeling community to standardise output. The foremost recommendation is that output is self-describing and unambiguous. This means adhering to the CF metadata conventions for representing time, including the use of CF boundary variables and cell method attributes. This explicitly distinguishes state variables from change variables and ensures that each variable has a stated reference time or period. 

We recommend that - in addition to total mass balance - individual mass-balance components be reported, which allows a more process-based model intercomparison. For example, identical annual mass balances may result from greater accumulation and greater melt in one model than in another. In addition, it is important that the two runoff definitions ('moving-gauge' and 'fixed-gauge') are kept clearly distinct from each other. Also, model-specific quantities and conventions, such as the value for ice density or the sign convention for mass-balance components, should be stated explicitly in the metadata. Importantly, none of these recommendations put constraints on model physics, calibration or initialisation. However, adopting them would considerably improve the ease of comparing published output, and make output more easily usable, such as in sea-level assessments or hydrological applications. \par 

Additionally, glacier model output, as well as the code and data required to produce it, should be openly archived in line with the FAIR principles \citep[Findable, Accessible, Interoperable and Reusable,][]{wilkinson2016fair}, which call for self-describing and standardised output. We recommend hosting output in a discipline-specific repository, ensuring quality control and long-term support. \par

Although the GlacierMIP4 protocol represents the most advanced standardization effort for global glacier modelling to date, its design still has several limitations. Because calibration is left to each participant, it is difficult to attribute inter-model differences to model structure or calibration. The reporting of the corrected forcing output only partially addresses this. Further refinement and updates will also be needed as glacier models evolve and resolve relevant processes in greater detail, allowing for more detailed output. For example, although the current protocol includes frontal ablation output, it does not distinguish between marine- and lake-terminating glaciers. Unlike ISMIP7, it also does not separate the individual components of frontal ablation, i.e., iceberg calving and ice front melt/sublimation. Any ablation, other than frontal ablation, is lumped into melt; sublimation is not considered separately. As global glacier models increasingly incorporate 2D or 3D ice-flow models, it will be valuable to include variables such as surface velocity and ice temperature, which are currently not part of the standard output.
Finally, as gridded global glacier models become more common, standardised gridded output fields will be useful.

In addition to expanding the set of output variables, advancing variable naming and unit conventions should be a high priority. Terminology should follow the definitions in \cite{cogley2011glossary}. Developing a common standard aligned as far as possible with ISMIP for a clearly defined set of standardised model output variables could be a direct objective for a future GlacierMIP effort. Variable names should adhere to the guidelines for construction of CF standard names and be incorporated into the CF standard name table, to improve consistency, interoperability, and clarity across glacier-model data products. 
Adhering to CF metadata conventions ensures consistency with ISMIP which uses these conventions to enable simulated ice-sheet variables to interact seamlessly with global climate model outputs - an important aspect given that ISMIP is integrated into the CMIP framework. Overall, establishing clear standards based on CF conventions is essential to ensure the traceability, interoperability, and comparability of global glacier projections.

The GlacierMIP4 protocol nonetheless represents an important first milestone toward standardising global glacier model output, and broad adoption of these conventions would substantially improve the comparability of glacier projections. As divergence in output conventions is a common obstacle in intercomparison efforts, also outside the narrow scope of global-scale glacier modelling, the standardisation recommendations outlined here may be similarly useful to the broader glacier modeling community. \\
\\

\section{ Supplementary Material} 
The GlacierMIP4 protocol can be found at \url{https://github.com/GlacierMIP/GlacierMIP4}. Supplementary Table S1 is included in the supplementary material. 

\section{ Acknowledgements} 
GlacierMIP is sponsored by the ‘Climate and Cryosphere’ (CliC) program of the World Climate Research Project (WCRP), and we thank in particular Meghan Taylor for her administrative support. JB, RH, RY and YAY received funding from the European Research Council (ERC) under the European Union’s Horizon Europe programme through the ERC-2022-ADG project GLACMASS (grant agreement No. 101096057). LS and HZ received funding from the European Research Council (ERC) under the European Union’s Horizon Framework research and innovation programme (grant agreement No 101115565; ‘ICE$^3$’ project). LVT and HZ were supported by an Odysseus Type II project (grant agreement no. G0DCA23N; “GlaciersMD” project. LvdL received funding from the Advanced Research and Invention Agency (ARIA) Forecasting Tipping Points programme, project G-PREWS (SCOP-PR01-P024) and acknowledges Horizon Europe funding under the ICELINK grant agreement Nº 101184621. LVT was supported by the European Space Agency projects Climate Space X-ECV “Karakoram” (contract nr 4000145271/24/I-LR). FM received support from the Research Software Maintenance Fund, managed by the Software Sustainability Institute and funded by UKRI grant reference AH/Z000114/1 and by the EU FP7 project LIQUIDICE (grant no. 101184962). Ibrar ul Hassan Akhtar provided comments on an earlier version of the protocol. 

\section{ Author Contribution Statement} 
RH and LvL led the project and wrote the paper with text input from YAY on GCM selection, and edits from all co-authors.
All except CW contributed to the development of the protocol, especially JB and RH.
YAY led the selection of the GCMs supported by analyses from LS. 
LvL led the compilation of Supplementary Table S1, with input from JB, RH, LvT, RY and YAY; MH, FM and DR reviewed and supplemented it.
JB implemented the protocol’s technical components.
RH and HZ co-chaired GlacierMIP.

\section{ Competing Interests} 
The authors declare none.


\bibliography{igsrefs}   
\bibliographystyle{igs}  

\end{document}


\nolinenumbers

\title[GlacierMIP Protocol]{Supplementary Material for 'The Glacier Model Intercomparison Project (GlacierMIP): Towards standardisation of global glacier projections'}

\author[van der Laan, Hock, and others]{Larissa Nora van der Laan$^{1*}$,
Regine Hock$^{2,3*}$, 
Johannes Brunner$^2$,
Yeliz A. Yilmaz$^2$,
Claire Wilson$^{9}$,
Lilian Schuster$^{10}$, 
Lander Van Tricht$^{5,6,10}$,
Ruitang Yang$^2$,
Harry Zekollari$^{10,5}$,
Matthias Huss$^{5,6}$, 
Ben Marzeion$^7$,
Fabien Maussion$^8$, 
David R. Rounce$^9$,
Johannes J. Fürst$^4$}

\affiliation{%
$^1$Niels Bohr Institute, Copenhagen University, Copenhagen, Denmark\\
$^2$Department of Geosciences, University of Oslo, Oslo, Norway\\
$^3$University of Alaska Fairbanks, Fairbanks, USA\\
$^4$Institute of Geography, Friedrich-Alexander-Universität Erlangen-Nürnberg, Erlangen, Germany\\
$^5$Laboratory of Hydraulics, Hydrology and Glaciology (VAW), ETH Z\"urich, Z\"urich, Switzerland\\
$^6$Swiss Federal Institute for Forest, Snow and Landscape Research (WSL), Sion, Switzerland\\
$^7$Institute of Geography and MARUM - Center for Marine Environmental Sciences, University of Bremen, Bremen, Germany\\
$^8$School of Geographical Sciences, University of Bristol, Bristol, UK\\
$^9$Department of Civil and Environmental Engineering, College of Engineering, Carnegie Mellon University, Pittsburgh, PA, USA\\
$^{10}$Department of Water and Climate, Vrije Universiteit Brussel, Brussel, Belgium\\[6pt]

  Correspondence: Regine Hock and Larissa Nora van der Laan 
  \email{regineho@uio.no, larissa.vdlaan@nbi.ku.dk}}

\begin{frontmatter}

\maketitle
\textsuperscript{*}~\textit{Contributed equally to this work.}
\end{frontmatter}
\clearpage

\setcounter{table}{0}
\renewcommand{\thetable}{S\arabic{table}}
\begin{longtable}{p{3cm} p{4.5cm} p{2.0cm} p{8.5cm}}

\caption{Non-exhaustive lookup table of output variables and conventions for the Python Glacier Evolution Model 
\cite[PyGEM v.1.1.0, \url{https://pygem.readthedocs.io/en/latest/},][]{Rounce_2023,tober_2026_projections}, the Open Global Glacier Model \cite[OGGM v.1.6.2, \url{https://oggm.org/},][]{Maussion2024_OGGM} and the Global Glacier Evolution Model 
\cite[GloGEM v.0.1.1, \url{https://glogem-docs.readthedocs.io/en/stable/index.html}][]{Huss_2015,Huss2026_GloGEM}.
"Variable" is the quantity being reported, named using each model's own terminology. "Short name" is the exact variable name used in each model's output file.}
\label{tab:lookup_S1} \\
\hline
\textbf{Variable} & \textbf{Short name} & \textbf{Unit} & \textbf{Definition / Notes} \\
\hline
\endfirsthead

\multicolumn{4}{l}{\tablename\ \thetable{} -- continued} \\
\hline
\textbf{Variable} & \textbf{Short name} & \textbf{Unit} & \textbf{Notes} \\
\hline
\endhead

\hline
\multicolumn{4}{l}{\textit{Continued on next page}} \\
\endfoot

\endlastfoot

\multicolumn{4}{l}{\textit{PyGEM}} \\
\hline

Glacier area & \texttt{glac\_area\_annual} & m$^2$ & Glacier area at start of the year \\

Glacier mass & \texttt{glac\_mass\_annual} & kg & Annual mass of ice based on area and ice thickness at start of the year (assuming ice density of 900 kg m$^{-3}$) \\

Mass below sea level & \texttt{glac\_mass\_bsl\_annual} & kg & Annual mass of ice below sea level, based on area and ice thickness at start of the year (assuming ice density of 900 kg m$^{-3}$ and sea level of 0 m a.s.l.) \\

Accumulation & \texttt{glac\_acc\_monthly} & m$^3$ w.e. & Solid precipitation only \\

Melt & \texttt{glac\_melt\_monthly} & m$^3$ w.e. & Melt of snow, firn and ice  \\

Frontal ablation & \texttt{glac\_frontalablation\_ monthly} & m$^3$ w.e. & Calving, subaerial and subaqueous frontal melt, and sublimation above the waterline. Positive values indicate mass loss. \\

Off-glacier runoff & \texttt{offglac\_runoff\_monthly} & m$^3$ & Runoff from area where the glacier no longer exists \\

Runoff & \texttt{glac\_runoff\_monthly} & m$^3$ & Runoff from the glacier area (melt-refreezing+rain) \\

Off-glacier runoff & \texttt{offglac\_runoff\_monthly} & m$^3$ & Runoff from initially glacierized area where the glacier no longer exists \\

Precipitation & \texttt{glac\_prec\_monthly} & m$^3$ & Liquid precipitation only, over glacier area\\

Time (nc-axis) & \texttt{time} & days since 2000-01-01 & Start of the month, no-leap calendar, calendar year \\

\hline

\multicolumn{4}{l}{\textit{OGGM\textsuperscript{a}}} \\
\hline

Glacier area & \texttt{area\_m2} & m$^2$ & Total glacier area at the start of the year \\

Glacier volume & \texttt{volume\_m3} & m$^3$ & Ice volume at the start of the year.\\

Glacier volume below sea level & \texttt{volume\_bsl\_m3} & m$^3$ & Ice volume below sea level (0 m a.s.l.) at the start of the year \\

Glacier mass & \texttt{mass\_kg} & kg & Glacier mass at the start of the year \\

Snowfall & \texttt{snowfall\_on\_glacier} & kg yr$^{-1}$ & Solid precipitation over the glacier area\\

Melt & \texttt{melt\_on\_glacier} & kg yr$^{-1}$ & Melt of snow, firn and ice over the glacier area, corrected so that snowfall minus melt matches the mass change of the dynamical model. Refreezing is not modelled. \\

Calving & \texttt{calving\_m3} & m$^3$ & Frontal ablation (calving) accumulated since the start of the run, in volume of ice\\

Liquid precipitation & \texttt{liq\_prcp\_on\_glacier} & kg yr$^{-1}$ & Rain over the glacier area (from precipitation after bias correction) \\

Time & \texttt{time} & year & Floating year, calendar years: \texttt{time} = $t$ is 1 January of year $t$. States (area, volume, mass, length) are given at the start of year $t$, hydrological fluxes are totals over year $t$. 365-day calendar by default.\\ 
\hline


\multicolumn{4}{l}{\textit{GloGEM}}\\
\hline

Glacier area & \texttt{area} & km$^2$ & Area of a glacier at the start of the year\\

Glacier volume & \texttt{vol} & km$^3$ & Volume of a glacier at the start of the year \\

Glacier volume below sea level & \texttt{SLE\_volbz} & km$^3$ & Volume of a glacier below sea level at the start of a year\\

Accumulation & \texttt{Accumulation\_sfc} & m w.e. yr$^{-1}$ & Annual accumulation over the glacier surface\\
Melt & \texttt{melt} & m w.e. yr$^{-1}$ & Annual glacier melt, separated into snowmelt and ice melt\\

Frontal ablation & \texttt{frontal\_ablation} & m w.e. yr$^{-1}$ & Glacier-wide specific mass loss due to frontal ablation at the terminus of an ocean-terminating glacier\\

Glacier runoff & \texttt{Discharge} & m w.e. yr$^{-1}$ & Glacier runoff over the evolving glacier area\\

Rain & \texttt{Rain} & m w.e. yr$^{-1}$ & Annual rainfall over the glacier surface\\

Time & \texttt{time} & year & Calendar year, values represent the end-of-year state \\
\hline


\end{longtable}
{\footnotesize
\textsuperscript{a}{Names as in the per-glacier output files (\texttt{model\_diagnostics*.nc}). In the regional files merged with \texttt{compile\_run\_output} (e.g.\ the OGGM standard projections), the unit suffixes \texttt{\_m3}, \texttt{\_m2}, \texttt{\_m} and \texttt{\_myr} are dropped (e.g.\ \texttt{volume}, \texttt{area})}}

\bibliography{igsrefs}   
\bibliographystyle{igs}